# Axionic Dark Matter Halos in the Gravitational Field of Baryonic Matter

Gennady P. Berman[1], Vyacheslav N. Gorshkov[2], and Vladimir I. Tsifrinovich[3],

[1]Theoretical Division, T-4, Los Alamos National Laboratory, Los Alamos, NM 87545, USA
[2]National Technical University of Ukraine "Igor Sikorsky Kyiv Polytechnic Institute", Kyiv, 03056, Ukraine
[3]Department of Applied Physics, NYU Tandon School of Engineering, Brooklyn, NY 11201, USA

**Abstract**

We consider a dark matter halo (DMH) of a spherical galaxy as a Bose-Einstein condensate (BEC) of the ultra-light axions (ULA) interacting with the baryonic matter. In the mean-field (MF) limit, we have derived the integro-differential equation of the Hartree-Fock type for the spherically symmetrical wave function of the DMH component. This equation includes two independent dimensionless parameters: (i) $\beta$ - the ratio of baryon and axion total masses and (ii) $\xi$ - the ratio of characteristic baryon and axion spatial parameters. We extended our "dissipation algorithm" for studying numerically the ground state of the axion halo in the gravitational field produced by the baryonic component. We calculated the characteristic size, $x_c$, of DMH as a function of $\beta$ and $\xi$. and obtained an analytical approximation for $x_c$.

## Introduction

Ultralight axions (ULA) have attracted much attention as a possible candidate for the dark matter halo (DMH) in galaxies (see, for example, [1-5]). Typical models suggest that axionic DMH contains a Bose-Einstein condensate (BEC) core and a quasi-classical axionic envelope (see, for example, [3,4]). In our previous work [5], we suggested that in some galaxies the whole DMH can be described by a simple model of the spherically symmetrical self-gravitating BEC completely ignoring the baryonic component. Certainly, it is possible only when the mass of the baryonic component is negligible compared to the mass of DMH.

It is known that in some galaxies the mass of the baryonic matter is essential. Moreover, recent observations show that the mass of the baryonic matter may exceed the mass of the DMH [6].

In this work we suggest that our simple model of spherically symmetrical axionic BEC [5] can be applied to the description of the DMH of some galaxies with the spherically symmetrical distribution of the baryonic mass for an arbitrary relation between the masses of the baryonic and axionic components. We consider a spherically symmetrical BEC of self-gravitating axions which experiences the gravitational field of a given spherically symmetrical distribution of the baryonic matter. We note that a closely related problem of axionic BEC, which experiences the gravitational field of a black hole, was considered in ref. [4]. In Section II, we derive, in the mean-field (MF) limit, the non-linear integro-differential equation for the axionic BEC in the presence of the baryonic component. In Section III, we describe our dissipation method for numerical solution of this equation, and in Section IV, we describe the results of our numerical simulations. The main objective of our paper is to calculate the characteristic size of DMH as a function of two independent parameters, the ratio of masses of the baryonic and axionic components and the size of the baryonic components. Our results may be relevant for description of recently discovered galaxies with the dominant baryonic component.

## II. Non-Linear Integro-Differential Equation for the axionic DMH in the presence of baryonic mass

We start with the non-relativistic Schrödinger equation,

$$i\hbar \partial \psi(\vec{r}_1,...,\vec{r}_N;t)/\partial t = H_N \psi(\vec{r}_1,...,\vec{r}_N;t), \tag{2.1}$$

for $N$ axions which interact gravitationally between themselves and with the given distribution of the baryonic matter. The system is described by the Hamiltonian,

$$H_N = \sum_{i=1}^{N} m_a U_b(\vec{r}_i) - \frac{\hbar^2}{2m_a}\sum_{i=1}^{N}\Delta_{\vec{r}_i} - Gm_a^2 \sum_{1\le i<j\le N}\frac{1}{|\vec{r}_i-\vec{r}_j|}. \tag{2.2}$$

In (2.2), $U_b(r)$ is a given baryonic potential,

$$U_b(r) = -GM_b \int \frac{\rho_b(r')d^3r'}{|\vec{r}-\vec{r}\,'|}, \tag{2.3}$$

where $m_a$ and $M_b$, are the mass of an axion and the total baryonic mass, correspondingly, and $\rho_b(r)$ is the probability density for the baryonic matter. We use the normalization condition for $\rho_b(r)$,

$$\int \rho_b(r)d^3r = 1, \tag{2.4}$$

For definiteness, we take the density of the baryonic matter for the spherical galaxies from ref. [7], and obtain the following normalized probability density,

$$\rho_b(r) = \frac{r_b}{4\pi r^2 (r+r_b)^2}, \qquad (2.5)$$

where $r_b$ is a characteristic dimensional parameter. Using, the approach, described in our previous work [5], we derive a single-particle equation of the Hartree-Fock type, in the mean-field (MF) limit,

$$i\hbar \frac{\partial}{\partial t}\psi(\vec{r},t) = \left[ -\frac{\hbar^2}{2m_a}\Delta_{\vec{r}} - Gm_a M_b \int \frac{\rho_b(r')d^3r'}{|\vec{r}-\vec{r}'|} - Gm_a M_a \int \frac{|\psi(\vec{r}',t)|^2 d^3r'}{|\vec{r}-\vec{r}'|} \right] \psi(\vec{r},t), \qquad (2.6)$$

Here $M_a = m_a N$ is the total mass of the axionic halo.

We will use the following notations,

$r_a = \dfrac{\hbar^2}{Gm_a^2 M_a}$ - characteristic size of the axionic halo with no baryons;

$\tau = \dfrac{\hbar}{m_a r_a^2} t$ - dimensionless time;

$\vec{x} = \dfrac{\vec{r}}{r_a}$ - dimensionless coordinate; $\qquad (2.7)$

$\phi(\vec{x},\tau) = r_a^{3/2} \psi(\vec{r},t)$ - dimensionless wave function;

$\rho_b(x) = r_a^3 \rho_b(r)$.

In the dimensionless notations (2.7), Eq. (6) becomes,

$$i\frac{\partial}{\partial \tau}\phi(\vec{x},\tau) = \left[ -\frac{1}{2}\Delta_{\vec{x}} - \beta \int \frac{\rho_b(\vec{x}')d^3\vec{x}'}{|\vec{x}-\vec{x}'|} - \int \frac{|\phi(\vec{x}',\tau)|^2 d^3\vec{x}'}{|\vec{x}-\vec{x}'|} \right] \phi(\vec{x},\tau),$$

$$\rho_b(x) = \frac{\xi}{4\pi x^2 (x+\xi)^2}, \qquad (2.8)$$

$$\int |\phi(\vec{x},\tau)|^2 d^3\vec{x} = 1, \quad \int \rho_b(x) d^3\vec{x} = 1.$$

In (2.8), two independent dimensionless parameters are introduced,

$$\beta = \frac{M_b}{M_a}, \; \xi = \frac{r_b}{r_a}, \qquad (2.9)$$

By using the explicit expression for the baryonic density, the baryonic potential in (2.8) becomes,

$$U_b(x) = -\beta \int \frac{\rho_b(x') d^3\vec{x}'}{|\vec{x}-\vec{x}'|} = -\xi\beta \left( \frac{1}{x}\int_0^x \frac{dx'}{(\xi+x')^2} + \int_x^\infty \frac{dx'}{x'(\xi+x')^2} \right) = -\frac{\beta}{\xi}\ln\left(1+\frac{\xi}{x}\right). \qquad (2.10)$$

### *Eigenvalue problem in the MF limit.*

Below, we are interested in the spherically symmetric ground-state eigenfunction, $\phi(x)$, of Eq. (2.8). The corresponding eigenvalue problem for Eq. (2.8) is described by the equation,

$$\varepsilon\phi(x) = \left( -\frac{1}{2}\Delta_x - \frac{\beta}{\xi}\ln\left(1+\frac{\xi}{x}\right) - \int \frac{|\phi(x')|^2 d^3\vec{x}'}{|\vec{x}-\vec{x}'|} \right)\phi(x),$$

$$\varepsilon = \frac{m_a r_a^2}{\hbar^2} E, \qquad (2.11)$$

where $E$ and $\varepsilon$ are the dimensional and dimensionless, eigenenergies of an axion, correspondingly. Eq. (2.11) can be rewritten as,

$$\varepsilon\phi(x) = -\frac{1}{2x}\frac{\partial^2}{\partial x^2}(x\phi(x)) - \frac{\beta}{\xi}\ln\left(1+\frac{\xi}{x}\right)\phi(x) - 4\pi\left( \frac{1}{x}\phi(x)\int_0^x s^2\phi^2(s) ds + \phi(x)\int_x^\infty s\phi^2(s) ds \right),$$

$$\phi(x=\infty) = 0, \; 4\pi\int_0^\infty x^2\phi^2(x) dx = 1. \qquad (2.12)$$

For numerical simulations, it is convenient to use the equation for,

$$\varphi(x) = \sqrt{4\pi}\, x\phi(x). \qquad (2.13)$$

We obtain from (2.12) the equations which include the effective potentials,

$$\varepsilon\varphi(x) = -\frac{1}{2}\frac{\partial^2\varphi(x)}{\partial x^2} + U(x)\varphi(x), \qquad (2.14)$$

$$U(x) = -\left(\frac{\beta}{\xi}\ln\left(1+\frac{\xi}{x}\right) + \frac{1}{x}\int_0^x \varphi^2(s)\,ds + \int_x^\infty \frac{\varphi^2(s)}{s}\,ds\right), \tag{2.15}$$

$$\varepsilon = \frac{1}{2}\int_0^\infty \left(\frac{\partial \varphi(x)}{\partial x}\right)^2 dx + \int_0^\infty U(x)\varphi^2(x)\,dx, \tag{2.16}$$

$$\varphi(0) = \varphi(\infty) = 0, \quad \int_0^\infty \varphi^2(x)\,dx = 1. \tag{2.17}$$

**III. Numerical Protocol**

The numerical solution of the system of equations (2.14) - (2.17) is based on an iterative method for finding a stationary solution to the equation,

$$\frac{\partial \varphi(x,t)}{\partial t} = \frac{i}{2}\frac{\partial^2 \varphi(x,t)}{\partial x^2} + i(\varepsilon(t) - U(x,t))\varphi(x,t), \tag{3.1}$$

in the region, $0 \leq x \leq X$. The value of the upper boundary of the computational domain, $X = 25$, guarantees a high accuracy of the solution to the problem, if in the integrals (2.14) - (2.17), it is used instead of infinity. One iteration cycle includes the following computational procedures. If the wave function, $\varphi^{(k)}(x,t)$, is known at the end of the performed iteration cycle with the number, $k$, then, based on this function, the potential, $U^{(k)}(x)$, and the energy, $\varepsilon^{(k)}$, are redefined (see Eqs. (2.15) and (2.16)). The calculated $U^{(k)}(x)$ and $\varepsilon^{(k)}$, are used in Eq. (3.1) for calculation of $\varphi^{(k+1)}(x, t^{(k)} + T)$ at the end of the $(k+1)$-th iteration cycle ($T$ is the iteration period). This procedure continues up to establishing of a stationary state.

Now, we discuss the important factors that provide fast convergence of iterations. The calculations of $\varphi^{(k+1)}(x,t)$, with the initial condition, $\varphi(x, t = t^{(k)}) = \varphi^{(k)}(x)$, are performed using the implicit difference method of the first order of accuracy in time [8],

$$\frac{\varphi_i^{j+1} - \varphi_i^j}{\Delta t} = \frac{i}{2}\frac{\varphi_{i+1}^{j+1} - 2\varphi_i^{j+1} + \varphi_{i-1}^{j+1}}{\Delta x^2} + i(\varepsilon^{(k)} - U^{(k)}(x))\varphi_i^{j+1}, \tag{3.2}$$

where $\Delta t$ and $\Delta x$ are the discretization steps in time and space, correspondingly. We used the values, $\Delta t = 0.5$ and $\Delta x = X/N$, $N = 4\times 10^4$. One iteration cycle assumes calculations of $J$ steps in time ($J = 300$, $T = J\Delta t$) in order to get the next approximation of the wave, $\varphi^{(k+1)}(x, t^{(k)} + T)$.

In the general case, this method of solving the Schrödinger equation is imperfect, since it leads to an unphysical attenuation of the wave function in time. However, in this iterative method, it is this attenuation that ensures the convergence of the proposed iteration method. Let us explain this statement in detail. Represent the wave function, $\varphi^{(k)}(x, t^{(k)})$, in the form of a series,

$$\varphi^{(k)}(x, t^{(k)}) = \sum a_n \hat{\varphi}_n(x), \qquad (3.3)$$

where $\hat{\varphi}_n(x)$ are exact solutions of Eq. (2.14) for energy levels, $\varepsilon_n$. The calculation of evolution of $\varphi^{(k)}(x,t)$, according to the scheme (3.3), leads to the expression [9],

$$\varphi^{(k)}(x,t) = \sum a_n \exp(i\delta E_n t) e^{-\gamma_n [t-t^{(k)}]} \hat{\varphi}_n(x), \quad \gamma_n = \delta E_n^2 \Delta t / 2, \quad \delta E_n = \varepsilon^{(k)} - \varepsilon_n. \qquad (3.4)$$

From (3.4), it is seen that the decomposition spectrum narrows over time and concentrates in the zone $\Delta \tilde{E}$, that covers the value, $\varepsilon^{(k)}$. Therefore, the calculation of the next approximate value of the energy, $\varepsilon^{(k+1)}$, by the function,

$\varphi^{(k+1)}(x,t) \equiv \varphi^{(k)}(x,t+T)$, on the $(k+1)$-th iteration-cycle, will give a value closer to one of the levels $\varepsilon_n \in \Delta \tilde{E}$ than on the $k$-cycle. As a result of a sequence of iterations, the solution of equation (3.1) converges to one of the exact solutions of Eq. (2.14): $\partial \varphi(x,t)/\partial t \to 0$ in (3.1). To which particular level - it depends on the initial approximation, chosen on the basis of qualitative physical considerations. Obviously, to search for the ground state, an iterative procedure should begin with a function of the form, $\varphi^{(0)}(x) = x\exp(-x^2/\delta^2)$. From this function, one should determine the potential, $U^{(0)}(x)$. The energy value, $\varepsilon^{(0)}$, should be set deliberately negative, since the relation (2.16) does not guarantee negativity of $\varepsilon^{(0)}$ when using the function, $\varphi^{(0)}(x)$, in it. In subsequent iterations, when calculating $\varepsilon^{(k)}$ and $U^{(k)}(x)$, it is preferable to use not the function $\varphi^{(k)}(x)$, but its module, $|\varphi^{(k)}(x)|$. If the solution is close to the exact one, then the real and imaginary parts of the wave function are similar to each other, so a replacing, $\varphi^{(k)}(x) \to |\varphi^{(k)}(x)|$, does not change the result. However, this technique accelerates the convergence of iterations to the exact solution.

Since the difference scheme (3.2) introduces an attenuation, the wave function should be renormalized to unity at the beginning of each iteration cycle according to (2.17). When approaching the stationary state, the corresponding normalization factor approaches unity, and the sequence of energy values, $\{\varepsilon^{(k)}\}$, also saturates. The control of these quantities determines the number of necessary iterations, $N_{iter}$.

In our problem, it is convenient to calculate the characteristics of the system on the plane $(\beta, \xi)$ along the trajectories, $\xi = \text{const}$, by changing the value, $\beta$. At the beginning of such trajectories ($\beta = 0$, baryons are absent), the wave function of axions is easily calculated in [5], and then it is used as an initial approximation for subsequent growth of $\beta$. Changing the parameter $\beta$ by 0.3 requires 7 iterations ($N_{iter} = 7$) to obtain a new solution (the error in the energy value, $|\delta\varepsilon| < 10^{-5}$, if we use the solution for the previous value of $\beta$ as the initial approximation).

In conclusion, we note that the use of the Crank and Nicolson difference scheme [9], for solving equation (3.1), does not lead to convergence of iterations, since such a scheme does not introduce attenuation into the wave function and does not narrow the expansion spectrum (3.4) from iteration to iteration.

## IV. Results

Our first objective was to calculate the characteristic dimensionless size, $x_c$, of the axionic DMH as a function of two dimensionless parameters: the ratio, $\beta$, of the total baryonic and axionic masses, and the size, $\xi$, of the baryonic component (in terms of the parameter, $r_a$). In [5], where the baryonic component was ignored $(\beta = 0)$, the size of the DMH was estimated under the condition that the probability distribution of the axionic component, $P(x) = |\phi(x)|^2$, approaches zero. Namely, $P(x_c = 8) \approx 0.0091 \times P(x = 0)$. With no baryonic component, according to [5], $x_c \approx 8$. The same criterion, for calculation of $x_c$, was used in this paper, when the baryonic component is present $(\beta \neq 0)$. In Fig. 1, we show the dependence of $x_c$ on two parameters, $\beta$ and $\xi$. As expected, the characteristic size of DMH drops with increase of the relative mass of the baryonic component, $\beta$, and decrease of its size, $\xi$. Fig. 1, also shows the eigenenergy of an axionic ground state, $\varepsilon$, as the function of the parameters, $\beta$ and $\xi$. In Fig. 2, we show the MF potential energy, $U(x)$, and the probability density, $P(x)$, for three concrete values of $\beta$ and $\xi$.

We have obtained an approximate analytical estimate for the characteristic size of the DMH, $x_c$, in terms of the corresponding size, $x_{c,0} \equiv x_c(\beta = 0)$, with no baryonic component,

$$x_c = x_{c,0} \times \left(1 + \frac{12\beta^{3/2}}{(1+2.5\beta)(1+\sqrt{\xi})}\right)^{-1}. \tag{4.1}$$

Fig. 3, demonstrates the validity of our analytical estimate (4.1). The solid lines represent the numerical solutions. The circles represent the approximation (4.1).

## V. Conclusion

In this paper, we follow our previous suggestion in ref. [5] that the DMH in some galaxies can be described as a BEC of ULA. We consider the spherical galaxies with an arbitrary relation between the masses of the baryonic and axionic components which may be especially relevant in connection to recently discovered galaxies with dominant baryonic component [6]. To take into consideration the baryonic component, we extended both our mathematical model based on MF limit and our numerical code based on the iterative approach. Using these techniques, we calculated the characteristic size of the DMH as a function of the ratio of masses of the baryonic and axionic components and characteristic size of the baryonic component. We have found an accurate analytical estimate for the DMH size. The results obtained in this paper may be relevant for the description of the DMH in galaxies with significant or dominant baryonic components.

## Acknowledgements

The work by G.P.B. was done at Los Alamos National Laboratory managed by Triad National Security, LLC, for the National Nuclear Security Administration of the U.S. Department of Energy under Contract No. 89233218CNA000001

**Figures**

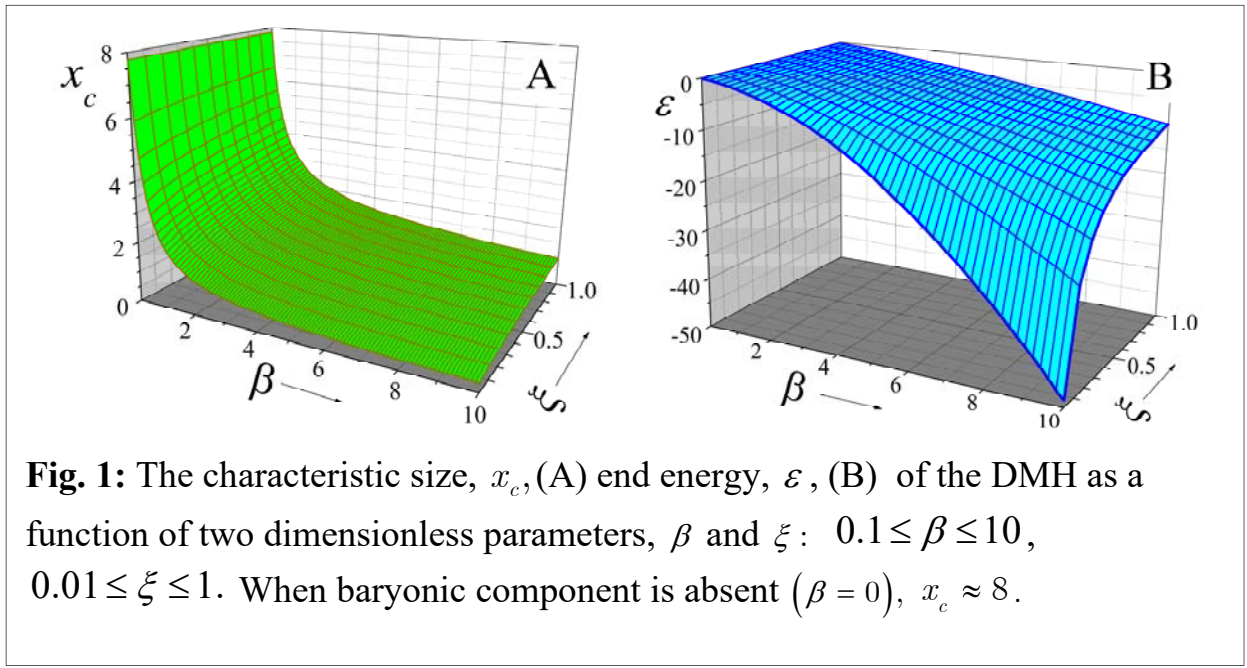

**Fig. 1:** The characteristic size, $x_c$, (A) end energy, $\varepsilon$, (B) of the DMH as a function of two dimensionless parameters, $\beta$ and $\xi$: $0.1 \leq \beta \leq 10$, $0.01 \leq \xi \leq 1$. When baryonic component is absent $(\beta = 0)$, $x_c \approx 8$.

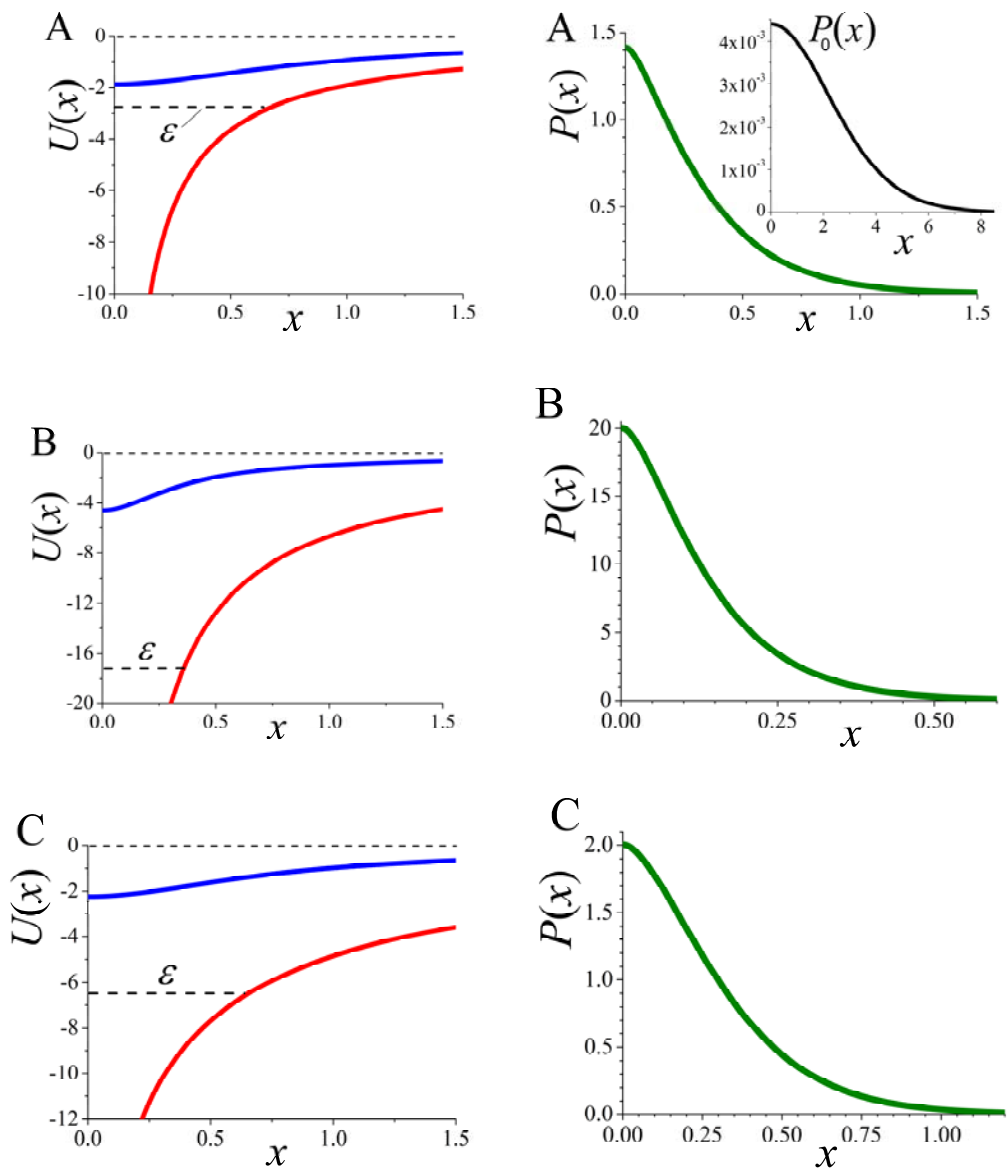

**Fig. 2:** Left – MF potential energy of DM axions (blue curves) and baryons (red curves). Right – probability function of axions, $P(x) = |\phi(x)|^2$. Upper insert – $P_0(x)$ - probability function of axions for $\beta = 0$, when baryons are absent ($\varepsilon \approx -0.1628$; the depth of the MF potential well - $U_0 \approx -0.3158$). **(A)** - $\xi = 0.1$, $\beta = 2$ **(B)** - $\xi = 0.1$, $\beta = 7$ **(C)** - $\xi = 1.0$, $\beta = 7$.

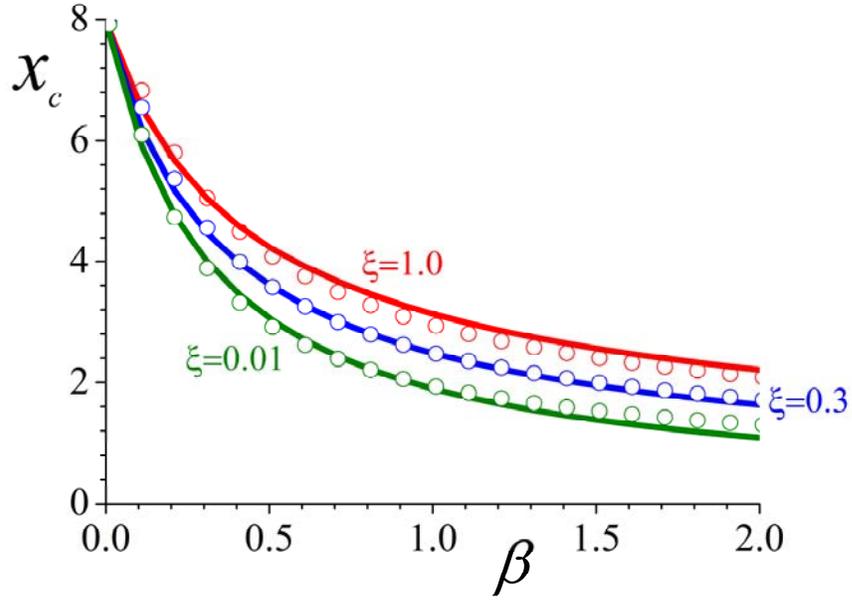

**Fig. 3:** The size of the DMH as a function of the ratio, $\beta$, of masses of baryonic and axionic components, at three values of the characteristic dimensionless size, $\xi$, of the baryonic density distribution.